# Computer simulated heat flow dynamics in silicon at low temperatures.

H. Diedrich*          R. Liberati**

The thermal characteristics of silicon between 15 and 300 deg K are investigated by applying a computer program on the solution of the differential heat diffusion equation. The computer model is linked to high-purity silicon through a set of experimental data. The numerical results are given in graphic form and show, in particular, very short diffusion transit times across long distances. The computed figures require experimental confrontations; a test set-up is proposed.

Pure silicon features particularly high values of both thermal conductivity and diffusivity as well as near-zero expansion coefficients at low temperatures, suggesting its use in heat transfer lines which bring the cold from the cryogenerator to the place where it is needed. This is rather problematic if a high-power heat flow has to be carried away from a distant or extended object. Silicon meets the power requirements under steady-state conditions. At LH2 temperatures however, where fast changes and in particular, first or second order phase transitions may occur, short dynamic response times, allowing for efficient feedback control mechanisms over long distances, are further needed. It was therefore the aim of this study to get deeper insight into the way temperature propagates across the silicon sample which in turn causes heat to flow down the temperature gradient into the cryogen.

This was done by simulating a "Thermal Discharge of a Silicon Heat Condenser", initially loaded with thermal energy at a temperature T(high), into a cryogen at T(low).

With some precautions the use of the phenomenological diffusion equation is reasonably (*) justified by the empirical evidence that heat diffusion between 300 and 15K is essentially ruled by phonon-phonon interaction, i.e. by the intrinsic thermal disorder of the crystal lattice continuum and that residual phonon scatter events at localised crystal imperfections of any kind, including surface scattering, create only marginal second order effects. The heat diffusion dynamics of this phonon continuum is characterised by the diffusivity D versus temperature T and is fully described by the appropriate solution of the diffusion equation.

The D(T) values which link the computer simulation to "real" silicon are calculated from published experimental data, listed on table 1, page 5 : the heat conductivities from 15 to 300 K had been measured on the same silicon slab with a boron concentration of 10^13 per ccm /1/ and some of them were taken from the interpolation curve "1M" in the original paper /2/. For the specific heat versus temperature, the tabulated values in /3/ were used. The thus calculated D versus T plot fits the calculated numerical function of Equ 3 over the entire 15 to 300 K range. The computer model is an "as-bulk" silicon rod which is thermally defined by Equ 3. Its linear dimension extends from x=0 to L, which is between 10 and 25 cm in the following examples. The left x(0) border is connected with the cryogen T(low) but is insulated from it by means of a "gate". At the time t(0) the gate starts opening with a speed, simulated by a "fast" or a "slow" Gaussian pulse, given by Equ 2.

---
(*) provided the phonon mean free path is small against the physical dimensions of the sample.

Because the heat diffusivity is temperature dependent the computer simulations are being performed by solving the nonlinear PDE /4/, /5/:

Equ 1a
$$c\left(x,t,u,\frac{\partial u}{\partial x}\right)\frac{\partial u}{\partial t} = \frac{\partial}{\partial x}\left(f\left(x,t,u,\frac{\partial u}{\partial x}\right)\right) + s\left(x,t,u,\frac{\partial u}{\partial x}\right)$$

where in our case are:   c = 1 and f = D(u(x,t)),    u(x,t) is the temperature field.

There is no heat source, the source term: $s\left(x,t,u,\frac{\partial u}{\partial x}\right)$ is zero. The solution components satisfy the initial condition of the form: $u(x,t_0) = u_0(x)$ and the boundary conditions are given by the following formula:

$$p(x,t,u) + q(x,t) \cdot f\left(x,t,u,\frac{\partial u}{\partial x}\right) = 0$$

The discretization in time and in 'x'-axis has been chosen in order to obtain a reasonably fine mesh and a calculation time of not more than one half hour on a Pentium IV with a single Intel processor of 2.0 GHz clock frequency.

The "gate" heat pulse has been simulated by

Equ 2          T(t) = ( T(h) -T(l) ) exp (- B t^2 ) + T(l)

         with          B = 1 / 2 sigma^2

The chosen sigma-values for fast and slow gates are  1/ sqrt (200) and  1/ sqrt (5), respectively.

The experimental D vs.T plot of table 1 is approximated by :

Equ 3              D(T) = A  exp (C ((lnT – B)^2) )

               with    A = 0.0008608
                       B = 10.3
                       C = 0.2955956252

The computer program uses therefore the following PDE

Equ 1b
$$\frac{\partial u}{\partial t} = \frac{\partial}{\partial x}\left(D(u) \cdot \frac{\partial u}{\partial x}\right)$$

together with the initial and boundary conditions as defined above. For the numerical calculation the temperature field u(x,t) is substituted by the experimental T-values of Equ 3.

The results of the computer simulation are given for the following examples with reference to figures 1 through 5.

( All numerical values are in cm / g / sec / deg K / watt units ).

| Example | high temp T(h) | low temp T(l) | length L | figures |
|---|---|---|---|---|
| 1 | 300 | 20 | 15 and 25 | 1: T vs (x,t)   2: dT/dx vs x   3: isotherms |
| 2 | 300 | 80 | 25 | 4: T vs (x,t) |
| 3 | 80 | 70 | 10 | 5: isotherms |
| 4 | 25 | 15 | 10 | --------- |

NB: The term "DT" in fig 2 stands for dT/dx

Example 1

The temperature in fig 1 shows a sharp edge below the 300K plateau and an almost abrupt fall of more than 200 degrees across a very thin spatial-temporal slice. (The apparent turbulence of the "breakdown" region is due to a poor space-time resolution capability of the computer program used). The phonon field fluidity increases with falling temperatures: The high temperature block "melts" away at the lowest T-values which causes the very narrow gradient peaks dT/dx, fig 2, with values of more than 1000 deg/cm, to move upwards from x=0 to L within a remarkably short time.

The T=constant projections on the (x,t) plane, fig 3, distinguish between an initial short near-left border zone and a linearly overlapping "bulk" zone in which the curves have quadratic forms. The isotherms fan out until the 300 K isotherm reaches the right border L. Then the thermal energy exhausts, the isotherms converge and the silicon heat condenser is cooling down until thermal equilibrium is reached. Due to the imposed operational bandwidth limit, the isotherms at lower T-values in fig 3 do not appear. The final cooling is anyway a slow process as it is shown by the Isotherms below.

Before the thermal energy starts exhausting, all isothermal curves have the general form

Equ 4 $$t = A\, x^2 + B$$

The value of A for the 300K isotherm in fig3 is approx. A = 1/300 sec/sqcm. By changing the gate speed from fast to slow, (not shown), the parameter B is increased from 0,26 to 1,25 sec, whereas the parameters A remain constant. The heat diffusion in the bulk is therefore not influenced by the gate condition.

Example 2

Temperature vs (x,t) depends strongly on the initial border temperatures, fig 4. Again the 300K isotherm (not shown) follows the parabolic law but A is much higher (1/18 sec/sqcm) and the arrival time at L is longer by a factor of about 20. After arrival the other isotherms change again their directions, as in the 20K case, from a quadratic function to higher dx/dt values. The B-value is small even at the slow gate condition.

Example 3

The fig 5 shows the heat discharge dynamics at lower temperatures and smaller high-low differences. The 80K isotherm follows the quadratic law, the transit time is 0,3 sec for L=10 cm and (not shown) increases with $L^2$. The heat flow down to 70K is slow, the border L reaches near-thermal equilibrium after a time which is one order of magnitude longer than the diffusion transit time.

Example 4

The diffusion dynamics at very low temperatures could only be estimated by extrapolation: In the particularly interesting liquid hydrogen range between triple point and boiling point the dT/dx peak travels across a 10 cm long rod in the order of milliseconds .

___________

Obviously, the picture presented by this study excludes surface effects which in reality will probably prevail over bulk properties at lower temperatures. In general, the computer simulation, though backed up by empirical data, is a mathematical exercise only and it will remain not more than that unless subjected to experimental trials (*): The applied heat condenser model indeed delivers some instructions for experimental checks. The dT/dx peak, though diffusion-broadened, creates a distinct dT/dt signal which should be measurable with a thermal sensor, integrated on the right L-border. The arrival time may then be defined by the condition: d/dt (dT/dt) = MIN!  for the double derivative of the sensor signal output.

______

(*) Appendix:  A test proposal.

A practical example of measuring the "diffusion transit time", is suggested by example 2:  In order to reduce surface scatter effects on the diffusion mechanism the test set-up operates at T(low) =77K,  in a range where phonon surface interactions play a minor role, where the B-component of Equ 4 is small and also, where the well-established LN2 technology  is helpful to supply simple technical options for the test set-up essentials. The cooling is performed by a high-pressure LN2 jet.  The "gate" is a screening shutter which is mechanically removed at t=0 within a very short time. The liquid-vapour jet, striking the silicon surface, causes an efficient turbulent heat exchange which may be estimated from fast JT-coolers data. The silicon test rods have limited surface/volume ratios, e.g. a diameter of 2cm and 5 to 10 cm lengths.  The surface is covered with a thermally grown silicon oxide layer, to be etched off in subsequent tests for evaluating different surface treatments. The cylindrical rod sits over its entire length in a vacuum chamber which is part of the provision to be made in order to satisfy the boundary conditions. The coaxial vacuum seal at the circumference near x=0 is performed by making use of the thermal match between silicon and some commercial glasses (see e.g. /4/). A low-inertia thermal sensor, e.g. in thin-film or in 2222-type version is applied on the L-face for measuring the temperature versus time profile in dc and in ac mode.

ADDRESSES

* Heinz Diedrich,    Vicolo Ciaccia 26,  00019 Tivoli (RM),  Italy
  Phone +39 0774 330069
  E-mail   tish@thehoys.com

** Riccardo Liberati,   Via della Rustica 277,  00155  Roma,   Italy
  Phone +39 0761 527136
  E-mail   siebrand@email.it

Edited July  2007

________________________________________________________________________________

TABLE 1

Experimental Data
Ref /1/ to /3/

| T (deg K) | k (watt / cm deg K) | c (watt sec / g deg K) | D *) (sq cm / sec) |
|---|---|---|---|
| 15  | 42    **) | 1,09  $10^{-3}$ | 16540 |
| 20  | 47,7       | 3,37  $10^{-3}$ | 6080  |
| 30  | 44,2       | 1,71  $10^{-2}$ | 1110  |
| 40  | 36,6       | 4,4   $10^{-2}$ | 357   |
| 50  | 28,0       | 7,9   $10^{-2}$ | 152   |
| 60  | 21         | 1,15  $10^{-1}$ | 78    |
| 80  | 12,5  **)  | 1,88  $10^{-1}$ | 29    |
| 100 | 9,13       | 2,59  $10^{-1}$ | 15,1  |
| 150 | 4,1        | 4,3   $10^{-1}$ | 4,0   |
| 200 | 2     **)  | 5,5   $10^{-1}$ | 1,53  |
| 300 | 0,85       | 6,9   $10^{-1}$ | 0,53  |

*) D = k / c d   with,  d = 2,33 g / ccm         **) interpolated (see "1M", /2/ )

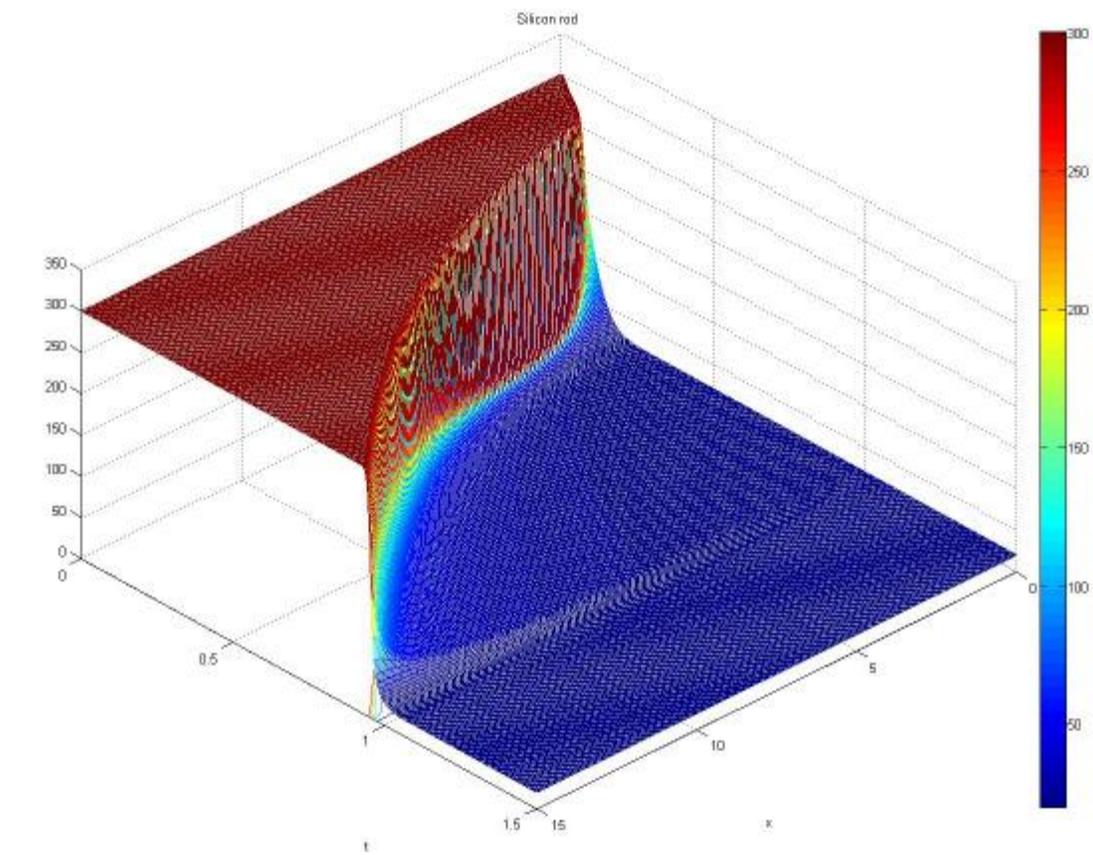
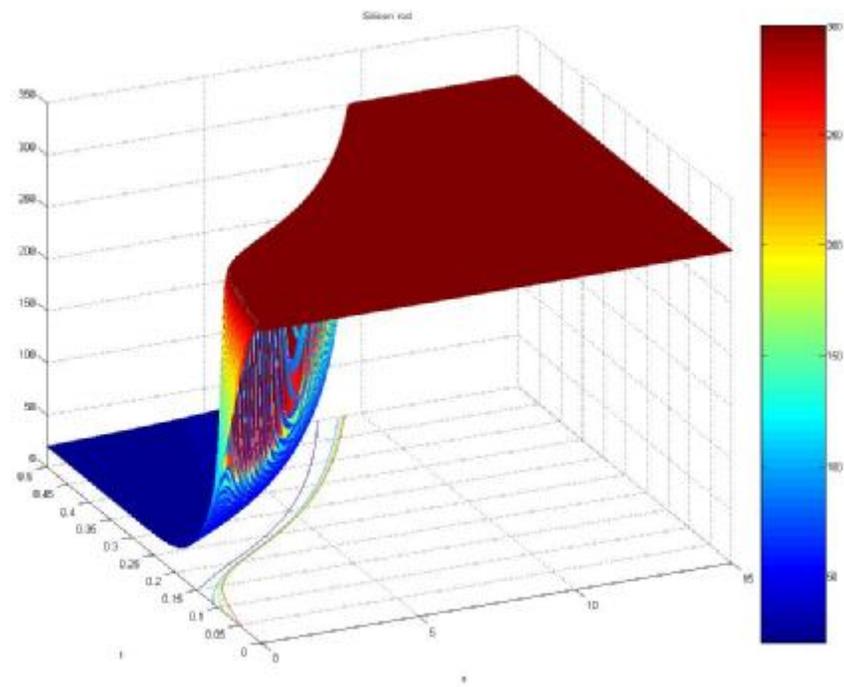

**Fig 1**

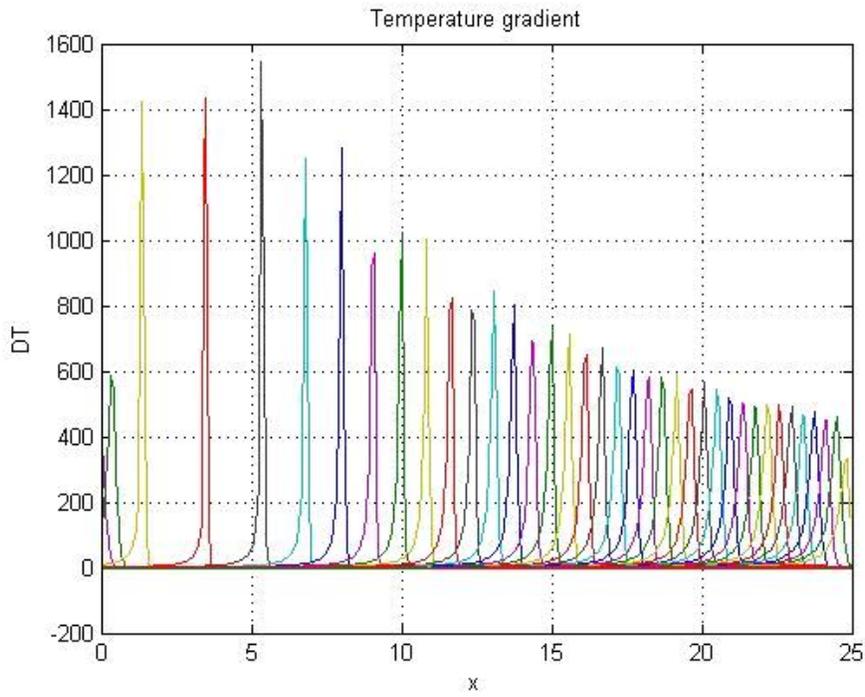

**Fig 2**

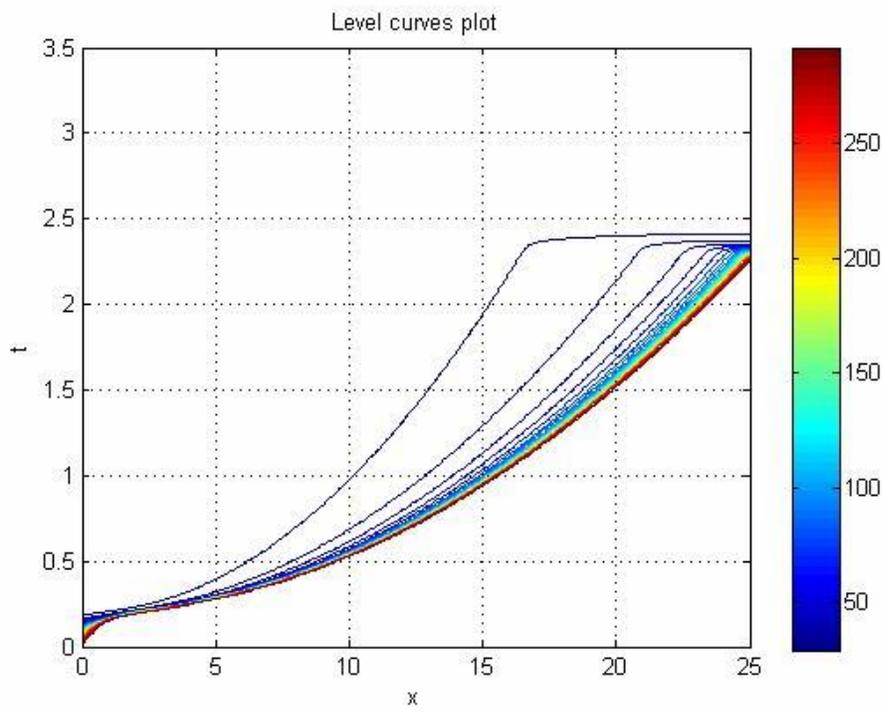

**Fig 3**

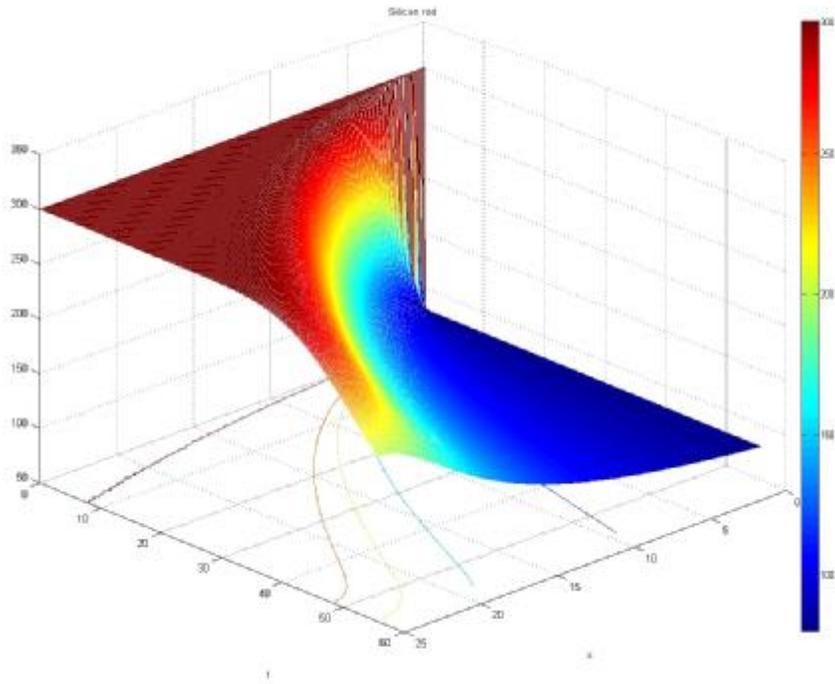

**Fig 4**

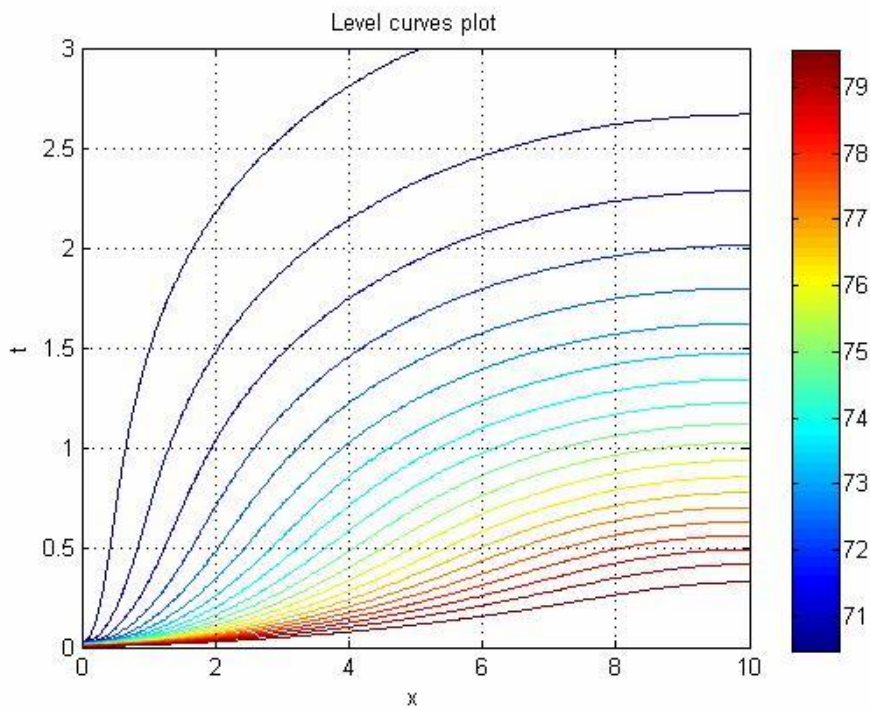

**Fig 5**